\documentclass[a4.11pt]{article}
\usepackage{color}
\usepackage{amssymb}
\usepackage{amsmath}
\usepackage{amsfonts}
\usepackage[mathscr]{eucal}
\usepackage{enumerate}
\usepackage{amscd}
\usepackage{indentfirst}
\usepackage{enumerate}
\usepackage{txfonts, textcomp}
\usepackage[dvips]{epsfig}

\def\<{\langle}
\def\>{\rangle}
\newtheorem{thm}{Theorem}[section]
\newtheorem{prop}[thm]{Proposition}
\newtheorem{dfn}[thm]{Definition}
\newtheorem{lem}[thm]{Lemma}
\newtheorem{ex}[thm]{Example}

\numberwithin{equation}{section}
\allowdisplaybreaks[0]

\begin{document}

%\markboth{Y. Hirota, D. Chru\'sci\'nski, T. Matsuoka, and M. Ohya}
%{On correlations and mutual entropy in quantum composite systems}

\title{\bf On correlations and mutual entropy \\ in quantum composite systems}

\author{Yuji HIROTA\\Department of Mathematics, Tokyo University of Science, Tokyo, 162-8601, Japan \and 
Dariusz CHRU\'SCI\'NSKI\\Institute of Physics, Nicolaus Copernicus University, Toru\'n, 87-100, Poland \and  
Takashi MATSUOKA\\Department of Business Administration and Information,\\ 
    Tokyo University of Science, Suwa, Nagano, 391-0292, Japan \and
and\\ Masanori OHYA\\Department of Information Science, Tokyo University of Science, Chiba, 278-8510, Japan}

\date{}
\maketitle

%\begin{history}
%\received{Day Month Year}
%\revised{Day Month Year}
%\accepted{Day Month Year}
%\comby{(xxxxxxxxxx)}
%\end{history}

\begin{abstract} We study the correlations of classical and quantum systems from the
information theoretical points of view.
 We analyze a simple measure of correlations based on entropy (such measure was already investigated
  as {\em the degree of entanglement} by Belavkin, Matsuoka and Ohya).
Contrary to naive expectation, it is shown that separable state might possesses stronger correlation
than an entangled state.
\end{abstract}

%%%%%%%%%%%%%%%%%%%%%%%%%%%%%%%%%%%%%%%%%%%%%%%%%%%

\section{Introduction}

Correlations play a key role both in classical and quantum physics. In
particular the study of correlations is crucial in many-body physics and
classical and quantum statistical physics. Recently, it turned out that
correlations play prominent role in quantum information theory and many
modern applications of quantum technologies and there are dozens of papers
dealing with this problem (for the recent review see e.g. \cite{HHHH}).

The aim of this paper is to analyze classical and quantum correlations
encoded in the bi-partite quantum states. Beside quantum entanglement we
analyze a new measure -- so called $D$-correlations -- and the quantum
discord. We propose to compare correlations of different bi-partite states
with the same reduces states, i.e. locally they contain the same
information. It is shown that surprisingly a separable state may be more
correlated that an entangled one. Analyzing simple examples of Bell diagonal
states we illustrate the behavior of various measures of correlations. We
also provide an introduction to bi-partite states and entanglement mappings
introduced by Belavkin and Ohya and recall basic notions from classical and
quantum information theory. An entanglement mapping encodes the entire
information about a bi-partite quantum state and hence it provides an
interesting way to deal with entanglement theory. Interestingly, it may be
applied in infinite-dimensional case and in the abstract $\mathbb{C}^*$%
-algebraic settings. Therefore, in a sense, it provides a universal tool in
entanglement theory.

The paper is organized as follows: in the next section we recall basic facts
from the theory of composite quantum systems and introduce the notion of
entanglement mappings. Moreover, we recall an interesting construction of
quantum conditional probability operators. Section \ref{E} recall classical
and quantum entropic quantities and collects basic facts from classical and
quantum information theory. In particular it contains the new measure of
correlation called $D$-correlation. Section \ref{DISCORD} recalls the notion
of \emph{quantum discord} which was intensively analyzed recently in the
literature. In section \ref{CIRC} we recall the notion of a circulant state
and provide several examples of states for which one is able to compute
various measures of correlations. Final conclusions are collected in the
last section.

Throughout the paper, we use standard notation: $\mathcal{H},\,\mathcal{K} $
for complex separable Hilbert spaces and denote the set of the bounded
operators and the set of all states on $\mathcal{H}$ by $\mathbf{B}(\mathcal{%
H})$ and $\mathbf{S}(\mathcal{H})$, respectively. In the $d$-dimensional
Hilbert space, the standard basis is denoted by $\{e_0,e_1,\cdots,e_{d-1}\}$
and the inner product is denoted by $\langle \cdot,\,\cdot\rangle$. We write
$e_{ij}$ for $|e_i\rangle\langle e_j|$. Given any state $\theta$ on the
tensor product Hilbert space $\mathcal{H}\otimes\mathcal{K}$, we denote by $%
\mathrm{Tr}_{\mathcal{K}}\theta$ the partial trace of $\theta$ with respect
to $\mathcal{K}$.

%%%%%%%%%%%%%%%%%%%%%%%%%%%%%%%%%%%%%%%%%%%%%%%%%%%%%%%%%%%%%%%%%%%%%%%%%%%%

\section{Quantum states and entanglement maps}

Consider a quantum system living in the Hilbert space $\mathcal{H}$. In this
paper we consider only finite dimensional case. However, as we shall see
several results may be nicely generalized to the infinite-dimensional
setting. Denote by ${\mathcal{T}(\mathcal{H})}$ a set of trace class
operators in $\mathcal{H}$, meaning that $\rho \in {\mathcal{T}(\mathcal{H})}
$ if $\rho \geq 0$ and $\mathrm{Tr}\, \rho < \infty$, which is always true
in finite-dimensional case. Finally, let
\begin{equation*}
{\mathbf{S}(\mathcal{H})} = \{\, \rho \in {\mathcal{T}(\mathcal{H})}\ |\
\mathrm{Tr}\, \rho =1\, \}\ ,
\end{equation*}
Consider now a composite system living in ${\mathcal{H} \otimes \mathcal{K}}$
and denote by $\mathbf{S}_{\mathrm{SEP}} \subset \mathbf{S}({\mathcal{H}
\otimes \mathcal{K}})$ a convex subset of separable states in ${\mathcal{H}
\otimes \mathcal{K}}$. Recall that $\rho \in \mathbf{S}({\mathcal{H} \otimes
\mathcal{K}})$ is separable if %\begin{equation}\label{}
$\rho = \sum_\alpha \, p_\alpha\, \eta_\alpha {\,\otimes\,} \sigma_\alpha$,
%\end{equation}
where $\eta_\alpha \in {\mathbf{S}(\mathcal{H})}$ and $\sigma_\alpha \in {%
\mathbf{S}(\mathcal{K})}$, and $p_\alpha$ denotes probability distribution: $%
p_\alpha \geq 0$ an $\sum_\alpha p_\alpha =1$. A state $\rho \in \mathbf{S}({%
\mathcal{H} \otimes \mathcal{K}})$ is called positive partial transpose
(PPT) if its partial transpose satisfies %\begin{equation}\label{}
$(\mathrm{id}_\mathcal{H} {\,\otimes\,} \tau)\rho \geq 0$, %\end{equation}
where $\mathrm{id}_\mathcal{H}$ denotes an identity map in ${\mathbf{B}(%
\mathcal{H})}$. It means that $\rho$ is PPT if $(\mathrm{id}_\mathcal{H} {%
\,\otimes\,} \tau)\rho \in \mathbf{S}({\mathcal{H} \otimes \mathcal{K}})$.
Denote by $\mathbf{S}_{\mathrm{PPT}}$ a convex subset of PPT states. It is
well known \cite{Pe} that %\begin{equation*}
$\mathbf{S}\left( \mathcal{H}\otimes\mathcal{K}\right)\supset \mathbf{S}_{%
\mathrm{PPT}}\supset \mathbf{S}_{\mathrm{SEP}}$. %\end{equation*}
In general, the PPT condition is not sufficient for separability.

Interestingly, due to the well known duality between states living in ${%
\mathcal{H} \otimes \mathcal{K}}$ and linear maps ${\mathbf{B}(\mathcal{K})}
{\, \rightarrow\, } {\mathbf{B}(\mathcal{H})}$, one may translate the above
setting in terms of linear maps. Let us recall basic facts concerning
completely positive maps \cite{Paulsen}. A linear map $\chi:\mathbf{B} (%
\mathcal{K}) \to \mathbf{B}(\mathcal{H})$ is said to be completely positive
(CP) if, for any $n\in\mathbb{N}$, the map
\begin{equation}
\chi _n: M_n(\mathbb{C})\otimes \mathbf{B}(\mathcal{K}) \longrightarrow
M_{n}(\mathbb{C})\otimes \mathbf{B}(\mathcal{H}), \quad (a_{i,j})_{i,j}
\longmapsto \bigl(\chi(a_{i,j})\bigr)_{i,j}
\end{equation}
is positive, where $\mathbf{B}(\mathcal{H})$ denotes bounded operators in $%
\mathcal{H}$ and $M_{n}(\mathbb{C})$ stands for $n\times n$ matrices with
entries in $\mathbb{C}$. A linear map $\chi:\mathbf{B}(\mathcal{K}) \to
\mathbf{B}(\mathcal{H})$ is said to be completely copositive (CCP) if
composed with transposition $\tau$, i.e. $\tau\circ\chi$, is CP.

Consider now a state $\theta \in \mathbf{S}({\mathcal{H} \otimes \mathcal{K}}%
)$ and let $\phi: \mathbf{B}(\mathcal{K})\to \mathbf{B}(\mathcal{H} )$ be a
linear map defined by
\begin{equation*}
\phi(b):= \mathrm{Tr}_{\mathcal{K}}\, [(1_\mathcal{H}\otimes b)\theta]\ ,
\end{equation*}
for any $b \in {\mathbf{B}(\mathcal{K})}$. The dual map $\phi^*$ reads
\begin{equation*}
\phi^*(a) = \mathrm{Tr}_{\mathcal{H}}\,[(a\otimes 1_\mathcal{K}) \theta] \ ,
\end{equation*}
for any $b \in {\mathbf{B}(\mathcal{H})}$. It should be stressed that the
above construction is perfectly well defined also in the
infinite-dimensional case if wew assume that $\theta$ is a normal state,
that is, it is represented by the density operator. Note, that a state $%
\theta$ and the linear map $\phi$ give rise a linear functional $\omega :
\mathbf{B}({\mathcal{H} \otimes \mathcal{K}}) {\, \rightarrow\, } \mathbb{C}$
\begin{equation}
\omega (a\otimes b) : = \mathrm{Tr}(a\otimes b) \theta,
\end{equation}
for any $a\in \mathbf{B}(\mathcal{H}),\,b\in \mathbf{B}(\mathcal{K})$. This
formula may be equivalently rewritten as follows
\begin{equation}
\omega (a\otimes b) =\mathrm{Tr}_{\mathcal{H}}~a\phi(b) = \mathrm{Tr}_{%
\mathcal{K}} ~ \phi^\ast(a) b.  \label{trace}
\end{equation}
It is clear that the marginal states read
\begin{equation}
\mathrm{Tr}_{\mathcal{K}}\theta = \phi(1_\mathcal{K})\in \mathbf{B}(\mathcal{%
\ H}),\quad \mathrm{Tr}_{\mathcal{H}}\theta =\phi^\ast(1_\mathcal{H})\in
\mathbf{B}(\mathcal{K}).
\end{equation}
Belavkin and Ohya observed \cite{BO1,BO2} that if $\theta \in \mathbf{S}({%
\mathcal{H} \otimes \mathcal{K}})$, then both $\phi$ and its dual $\phi^*$
are CCP. We denote by $\mathbf{B}\mathcal{(H)}$ the dual space to the
algebra $\mathbf{B}\mathcal{(H)}$.

\begin{dfn}
A CCP map $\phi:\mathbf{B}\mathcal{(K)}\rightarrow\mathbf{B}\mathcal{(H)} $
normalized as $\mathrm{Tr}_{\mathcal{H}}\phi (1_{\mathcal{K}}) =1$ is called
the entanglement map from $\rho:=\phi ^{\ast}(1_{\mathcal{H}})\in \mathbf{B}%
\mathcal{(K)}$ to $\sigma:=\phi (1_{\mathcal{K}})\in \mathbf{B}(\mathcal{H}
) $.
\end{dfn}

A density operator $\theta _{\phi }$ corresponding to the entanglement map $%
\phi$ with its marginals $\phi ^{\ast}(1_{\mathcal{H}})$ and $\phi (1_{
\mathcal{K}})$ can be represented as follows: let $\psi^+_\mathcal{K}$
denotes a maximally entangled state in $\mathcal{K}{\,\otimes\,} \mathcal{K}$%
. Then
\begin{equation}
\theta_\phi := (\phi {\,\otimes\,} \tau) P^+_\mathcal{K}\ ,
\end{equation}
with $P^+_\mathcal{K} = d_\mathcal{K}\,|\psi^+_\mathcal{K}\>\<\psi^+_%
\mathcal{K}|$, where $d_\mathcal{K} = \mathrm{dim}\,\mathcal{K}$. If $\{ e_k
\}$ stands for an orthonormal basis in $\mathcal{K}$, then
\begin{equation}
P^+_\mathcal{K} = \sum_{i,j=1}^{d_\mathcal{K}} e_{ij} {\,\otimes\,} e_{ij}\ ,
\end{equation}
with $e_{ij}:= |e_i\>\<e_j|$, and hence
\begin{equation}
\theta_{\phi}= \sum_{i,j=1}^{d_\mathcal{K}}\, \phi(e_{ji}) \otimes e_{ij} \ .
\label{def1}
\end{equation}
The map assigning $\theta_\phi$ to $\phi$ is usually called a Choi-Jamio\l %
kowski isomorphism. It should be stressed that $\theta_\phi$ does not depend
upon the choice of $\{e_k\}$.

\begin{lem}
\label{w-criterion} A linear map $\phi : {\mathbf{B}(\mathcal{K})}%
\rightarrow {\mathbf{B}(\mathcal{H})}$ is CCP if and only if $\theta _{\phi
}\geq 0$. Clearly, $\phi $ is CP if and only if $\phi \circ \tau $ is CCP.
\end{lem}

Due to Lemma \ref{w-criterion}, we have the following criterion.

\begin{thm}
\cite{JMO,MMO} A state $\theta _{\phi }$ is a PPT state if and only if its
entanglement map $\phi $ is CP.
\end{thm}

Recently, Kossakowski et al.\cite{AKMS} proposed the following construction:
for $\theta \in \mathbf{S}(\mathcal{H} \otimes\mathcal{K})$ one defines the
bounded operator
\begin{equation}
\pi_{\theta}:= \bigl(\rho^{-\frac{1}{2}}\otimes 1_\mathcal{K}\bigr)\,
\theta\,\bigl(\rho^{-\frac{1}{2}}\otimes 1_\mathcal{K}\bigr),
\end{equation}
where $\rho :=\mathrm{Tr}_{\mathcal{K}}\theta$. It is verified that $%
\pi_\theta$ satisfies
\begin{align}
\pi_{\theta }&\geq 0,  \label{con1} \\
\mathrm{Tr}_{\mathcal{K}}\pi_\theta &= 1_\mathcal{H} \in \mathbf{B}(\mathcal{%
\ H}).  \label{con2}
\end{align}
In what follows we assume that $\rho$ is a faithful state, i.e. $\rho >0$.
It follows from ( \ref{con1}) and (\ref{con2}) that the operator $\pi_\theta$
is the quantum analogue of a classical conditional probability. Indeed, if $%
\mathbf{B}( \mathcal{H}\otimes\mathcal{K})$ is replaced by commutative
algebra, then $\pi_\theta$ coincides with a classical conditional
probability.

\begin{dfn}
An operator $\pi\in \mathbf{B}(\mathcal{H}\otimes\mathcal{K})$ is called the
quantum conditional probability operator (QCPO, for short) if $\pi$
satisfies condition (\ref{con1}) and (\ref{con2}).
\end{dfn}

It is easy to verify\cite{AKMS} that for any CP unital map $\varphi : {%
\mathbf{B}(\mathcal{K})} {\, \rightarrow\, }{\mathbf{B}(\mathcal{H})}$ and
an orthonormal basis in $\mathcal{K}$ the following operator
\begin{equation}
\pi_\varphi=\sum_{k,l=1}^{d_\mathcal{K}}\varphi(e_{kl}) \otimes e_{kl} \ ,
\label{def2}
\end{equation}
defines QCPO. From Lemma \ref{w-criterion} and unitality of $\varphi$, it
follows that $\pi _{\varphi }$ satisfies conditions (\ref{con1}) and (\ref%
{con2}). For a given $\pi_{\varphi }$ and any faithful marginal state $\rho
\in \mathbf{S}(\mathcal{H })$, one can construct a state $\theta$ of the
composite system
\begin{equation}
\theta_\varphi = \sum_{k,l=1}^{d_\mathcal{K}} \rho^{\frac{1}{2}%
}\,\varphi(e_{kl}) \rho^{\frac{1}{2}}\otimes e_{kl}\ .
\end{equation}
It is clear that $\theta_\varphi$ is a PPT state if and only if the map $%
\varphi$ is a CCP. There exists a simple relation between the density
operator $\theta_\phi$ in (\ref{def1}) and the QCPO $\pi_\varphi$ in (\ref%
{def2}) due to the following decomposition of the entanglement map $\phi $.

\begin{lem}
\cite{BD} Every entanglement map $\phi $ with $\phi(1_\mathcal{K}) =\rho$
has a decomposition
\begin{equation}
\phi \left( \cdot \right) =\rho ^{\frac{1}{2}}\varphi \circ \tau \left(
\cdot \right) \rho ^{\frac{1}{2}},
\end{equation}
where $\varphi $ is a CP unital map to be found as a unique solution to
\begin{equation}
\varphi(\cdot ) =\rho ^{-\frac{1}{2}}\phi \circ \tau(\cdot) \rho ^{-\frac{1}{
2}}.
\end{equation}
\end{lem}

\begin{thm}
\cite{CKMO} If a composite state $\theta _{\phi }$ given by (\ref{def1}) has
a faithful marginal state $\rho =\phi(1_\mathcal{K})$, then $\theta_\phi$ is
represented by
\begin{equation}
\theta_\phi=\bigl(\rho^{\frac{1}{2}}\otimes 1_\mathcal{K}\bigr)\,
\pi_{\phi}\,\bigl(\rho^{\frac{1}{2}}\otimes 1_\mathcal{K}\bigr),
\label{FundaQ}
\end{equation}
where $\pi_\phi=\sum_{k,l}\rho^{-\frac{1}{2}}\,\phi(e_{kl})\,\rho^{-\frac{1}{%
2}}\otimes e_{kl}$.
\end{thm}

%%%%%%%%%%%%%%%%%%%%%%%%%%%%%%%%%%%%%%%%%%%%%%%%%%%%%%%%%%%%%%%%%%%%%%%%%%%%%%%%%

\section{Classical and quantum information}

\label{E}

In classical description of a physical composite system its correlation can
be represented by a joint probability measure or a conditional probability
measure. In classical information theory we have proper criteria to estimate
such correlation, which are so-called the mutual entropy and the conditional
entropy given by Shannon \cite{Sh}. Here we review Shannon's entropies
briefly.

Let $X=\{x_{i}\}_{i=1}^{n}$ and $Y=\{y_{j}\}_{j=1}^{m}$ be random variables
with probability distributions $p_{i}$ and $q_{j}$, respectively, and let $%
p_{i|j}$ denotes conditional probability $P(X=x_{i}|Y=y_{j})$. The joint
probability $r_{ij}=P(X=x_{i},Y=y_{j})$ is given by
\begin{equation}
r_{ij}=p_{i|j}\,q_{j}\ .  \label{FundaC}
\end{equation}%
Let us recall definitions of mutual entropy $I(X:Y)$ and conditional
entropies $S(X\mid Y),\,S(Y\mid X)$:
\begin{equation*}
I(X:Y)=\sum_{i,j}\,r_{ij}\log \frac{r_{ij}}{p_{i}q_{j}}\ ,
\end{equation*}%
and
\begin{equation*}
S(X\mid Y)=-\sum_{j}q_{j}\sum_{i}p_{i|j}\log p_{i|j}\ ,\ \ \ \ \ S(Y\mid
X)=-\sum_{i}p_{i}\sum_{j}p_{j|i}\log p_{j|i}\ .
\end{equation*}
Using (\ref{FundaC}), we can easily check that the following relations
\begin{equation}
I(X:Y)=S(X)+S(Y)-S(XY)\ , \label{relation1}
\end{equation}
and
\begin{align}
S(X\mid Y)& =S(XY)-S(Y)=S(X)-I(X:Y)\ ,  \label{relation2} \\
S(Y\mid X)& =S(XY)-S(X)=S(Y)-I(X:Y)\ ,  \label{relation3}
\end{align}
where $S(X)=-\sum_{i}p_{i}\log p_{i}\,$, and $S(XY)=-\sum_{ij}r_{ij}\log r_{ij}$. 
Note, that $p_{i|j}$ gives rise to a stochastic matrix $T_{ij}:=p_{i|j}$ and hence it defines a classical channel
\begin{equation}
p_{i}=\sum_{j}T_{ij}q_{j}\ .
\end{equation}
Note, that data provided by $r_{ij}$ are the same as those provided by $T_{ij}$ and $p_{j}$. Hence one may instead of $I(X:Y)$ use the following
notation $I(P,T)$, where $P$ represent an \emph{input} state and $T$ the
classical channel. One interprets $I(P,T)$ as a information transmitted
\emph{via} a channel $T$. The fundamental Shannon inequality
\begin{equation}
0\leq I(P;T)\leq \min \bigl\{S(X),\,S(Y)\bigr\}\ ,  \label{FundaI}
\end{equation}
gives the obvious bounds upon the transmitted information.

Now, we extend the classical mutual entropy to the quantum system using the
Umegaki relative entropy.\cite{Ume} Let $\theta \in \mathbf{S}({\mathcal{H}
\otimes \mathcal{K}})$ with marginal states $\rho \in \mathbf{S}(\mathcal{H}) $ and $\sigma\in \mathbf{S}(\mathcal{K})$. One defines quantum mutual
entropy as a relative entropy between $\theta$ and the product of marginals $\rho {\,\otimes\,} \sigma$:
\begin{equation}
I(\theta) = S(\theta\,||\,\rho {\,\otimes\,} \sigma) = \mathrm{Tr}\, \{
\theta\bigl(\log \theta - \log [\rho \otimes\sigma] \bigr) \}\ .
\end{equation}
As in the classical case one shows that
\begin{equation}
I(\theta) = S(\rho) + S(\sigma) - S(\theta)\ .
\end{equation}
Introducing quantum conditional entropy
\begin{equation}
S_\theta(\rho\,|\,\sigma) := S(\theta) - S(\sigma)\ ,
\end{equation}
one finds
\begin{equation}
I(\theta) = S(\rho) - S_\theta(\rho\,|\,\sigma)\ ,
\end{equation}
or, equivalently
\begin{equation}
I(\theta) = S(\sigma) - S_\theta(\sigma\,|\,\rho)\ .
\end{equation}

\begin{dfn}
\cite{BO1,BO2,Cerf,Gro} For any entanglement map $\phi:\mathbf{B}(\mathcal{K}%
)\to\mathbf{B}(\mathcal{H})$ with $\rho = \phi(1_\mathcal{K})$ and $\sigma
=\phi^\ast(1_\mathcal{H})$, the quantum mutual entropy $I_\phi(\rho :\sigma)$
is defined by
\begin{equation}
I_\phi(\rho :\sigma) := S(\theta_\phi\, ||\, \rho \otimes \sigma) = \mathrm{
Tr}\, \{\theta_\phi\bigl(\log \theta _\phi - \log [\rho \otimes\sigma] 
\bigr) \}\ ,
\end{equation}
where $S(\cdot\, ||\, \cdot) $ is the Umegaki relative entropy.
\end{dfn}

One easily finds
\begin{align}
I_\phi(\rho : \sigma) = S(\rho) + S(\sigma) -S(\theta_\phi)\ .
\label{relation4}
\end{align}
The above relation (\ref{relation4}) is a quantum analog of (\ref{relation1}). 
One defines the quantum conditional entropies as generalizations of (\ref{relation2}), (\ref{relation3}) \cite{BO1,BO2,Cerf,HH}:
\begin{align}
S_\phi(\sigma\,|\,\rho) := S(\sigma) -I_\phi(\rho :\sigma) =S(\theta_\phi) -
S(\rho)\ .
\end{align}
It is usually assumed that $I_\phi(\rho:\sigma)$ measures all correlations
encoded into the bipartite state $\theta_\phi$ with marginals $\rho$ and $%
\sigma$.

\begin{ex}[Product state]
For the entanglement map
\begin{equation*}
\phi(b) := \rho\, \mathrm{Tr}_\mathcal{K}(\sigma b)\ ,
\end{equation*}
one finds $\theta _{\phi }=\rho \otimes \sigma$, and hence
\begin{align}
I_\phi(\rho : \sigma) = 0 \ , \ \ S_{\theta }\bigl(\sigma\,|\,\rho \bigr) =
S(\sigma)\ , \ \ S_{\theta }\bigl(\rho\,|\,\sigma \bigr) = S(\rho) \ ,
\end{align}
which recover well known relations for a product state $\rho {\,\otimes\,}
\sigma$.
\end{ex}

\begin{ex}[Pure entangled state]
\label{example:pure entangled} Let $\{\lambda _{i}\}$ be the sequence of
complex numbers satisfying $\sum_{i}|\lambda _{i}|^{2}=1$. For entanglement
mappings
\begin{equation*}
\phi (b)=\sum_{i,j=1}^{r}\, \lambda _{i}\overline{\lambda }_{j}\, e_{ij}\,
\langle f_{j},\,bf_{i}\rangle \ ,
\end{equation*}
where $\{ e_k\}$ and $\{f_l\}$ are orthonormal basis in $\mathcal{H}$ and $%
\mathcal{K}$, respectively, the state $\theta _{\phi }$ can be written in
the following form
\begin{equation*}
\theta _{\phi }=\sum_{i,j=1}^{r}\lambda _{i}\,\overline{\lambda }_{j}\,
e_{ij} \otimes f_{ij} =\bigl\vert\Psi \rangle \langle \Psi \bigr\vert\ ,
\end{equation*}
where
\begin{equation*}
\bigl\vert\Psi \bigr\rangle=\sum_{i=1}^r\lambda _{i}\,e_{i}\otimes f_{i} \in
{\mathcal{H} \otimes \mathcal{K}}\ .
\end{equation*}
Note, that
\begin{equation*}
r \leq \min\{ d_\mathcal{H},d_\mathcal{K}\} \ ,
\end{equation*}
equals to the Schmidt rank of $\Psi \in {\mathcal{H} \otimes \mathcal{K}}$.
One finds for the reduced states
\begin{equation*}
\rho =\phi (1_{\mathcal{K}})=\sum_{i=1}^r|\lambda _{i}|^{2} e_{ii}\ , \ \ \
\ \sigma =\phi ^{\ast }(1_{ \mathcal{H}})=\sum_{i=1}^r|\lambda _{i}|^{2}
f_{ii}\ ,
\end{equation*}
and hence
\begin{align}
I_\phi(\rho : \sigma) =S(\rho )+S(\sigma )-S(\theta) =2S(\rho )>\min \bigl\{%
S(\rho ),S(\sigma )\bigr\}\ ,
\end{align}
together with
\begin{align}
S_{\theta}(\sigma | \rho ) =S_{\theta }(\rho | \sigma )=-S(\rho )<0,
\end{align}
where $S(\rho )=S(\sigma )=-\sum_{i=1}^r|\lambda _{i}|^{2}\log |\lambda
_{i}|^{2} $.
\end{ex}

As is mentioned in Section 2, the classical mutual entropy always satisfies
the Shannon's fundamental inequality, i.e. it is always smaller than its
marginal entropies, and the conditional entropy is always positive. Note
that separable state has the same property. It is no longer true for pure
entangled states.

Now we introduce another measure for correlation of composite states.\cite%
{BO1,BO2,CKMO,MO}

\begin{dfn}
For the entanglement map $\phi: \mathbf{B}(\mathcal{K})\to\mathbf{B}(%
\mathcal{H})$, we define the $D$-correlation $D(\theta)$ of $\theta$ as
\begin{align}
D(\theta) &:= -\frac{1}{2}\left\{ S_{\theta}(\sigma|\rho) + S_\theta(\rho|
\sigma)\right\} = \frac 12 ( S(\rho) + S(\sigma)) - S(\theta) \ .
\label{DEN}
\end{align}
\end{dfn}

Note that the $D$-correlation with the opposite convention $-D(\theta)$ is
called the degree of entanglement.\cite{BO1,BO2,CKMO,MO} One proves the
following:

\begin{prop}
\label{section4:purecase} \cite{AMO,MO} If $\theta_\phi$ is a pure state,
then the following statements hold:

\begin{enumerate}
\item \textrm{$\theta$ is entangled state if and only if $D(\theta) >0$. }

\item \textrm{$\theta $ is separable state if and only if $D(\theta) =0$. }
\end{enumerate}
\end{prop}

It is well-known that if $\theta$ is a PPT state, then
\begin{equation}
S(\theta) -S(\rho) \geq 0,\quad S(\theta)-S(\sigma) \geq 0,
\end{equation}
where $\rho $ and $\sigma $ are the marginal states of $\theta$.\cite{VW}

\begin{prop}
\label{section4:mixedcase} If $\theta$ is a PPT state, then
\begin{equation}
D(\theta)\leq 0.
\end{equation}
\end{prop}

Suppose now that we have two entanglement mappings $\phi_k:\mathbf{B}(%
\mathcal{K})\to\mathbf{B}(\mathcal{H}),\,(k=1,2)$ such that $\phi_1(1_%
\mathcal{K})=\phi_2(1_\mathcal{K})$ and $\phi_1^\ast(1_\mathcal{H}%
)=\phi_2^\ast(1_\mathcal{H})$. Let $\theta_1, \theta_2 \in \mathbf{S}({%
\mathcal{H} \otimes \mathcal{K}})$ be the corresponding states. We propose
the following:

\begin{dfn}
$\theta_1$ is said to have stronger $D$-correlations than $\theta_2$ if
\begin{equation}
D(\theta_1) > D(\theta_2)\ .  \label{order}
\end{equation}
\end{dfn}

Several measures of correlation based on entropic quantities were already
discussed by Cerf and Adami\cite{Cerf}, Horodecki\cite{HH}, Henderson and
Vedral\cite{Vedral}, Groisman et al.\cite{Gro}.

\section{Quantum discord}

\label{DISCORD}

Let us briefly recall the definition of quantum discord \cite{Zurek,Vedral}.
Recall, that mutual information may be rewritten as follows
\begin{equation}
\mathcal{I}(\theta) = S(\sigma) - S_\theta(\sigma|\rho) \ .
\end{equation}
An alternative way to compute the conditional entropy $S_\theta(\sigma|\rho)$
goes as follows: one introduces a measurement on $\mathcal{H}$-party defined
by the collection of one-dimensional projectors $\{\Pi_k\}$ in $\mathcal{H}$
satisfying $\Pi_1 + \Pi_2 + \ldots = 1_\mathcal{H}$. The label `$k$'
distinguishes different outcomes of this measurement. The state after the
measurement when the outcome corresponding to $\Pi_k$ has been detected is
given by
\begin{equation}
\theta_{\mathcal{K}|k} = \frac{1}{p_k} (\Pi_k {\,\otimes\,} 1_\mathcal{K}%
)\theta (\Pi_k {\,\otimes\,} 1_\mathcal{K})\ ,
\end{equation}
where $p_k$ %$p_k = {\rm tr}[\rho_{B|k} (\Pi_k\ot \mathbb{I}_B)]$
is a probability that $\mathcal{H}$-party observes $k$th result, i.e. $p_k =
\mathrm{Tr}(\Pi_k \rho)$, and $\theta_{\mathcal{K}|k}$ is the (collapsed)
state in ${\mathcal{H} \otimes \mathcal{K}}$, after $\mathcal{H}$-party has
observed $k$th result in her measurement. The entropies $S(\theta_{\mathcal{K%
}|k})$ weighted by probabilities $p_k$ yield the conditional entropy of part
$\mathcal{K}$ given the complete measurement $\{\Pi_k\}$ on the part $%
\mathcal{H}$
\begin{equation}
S(\theta|\{\Pi_k\}) = \sum_k p_k S(\theta_{\mathcal{K}|k})\ .
\end{equation}
Finally, let
\begin{equation}
\mathcal{I}(\theta|\{\Pi_k\}) = S(\sigma) - S(\theta|\{\Pi_k\}) \ ,
\end{equation}
be the corresponding measurement induced mutual information. The quantity
\begin{equation}  \label{C-sup}
\mathcal{C}_{\mathcal{H}}(\theta) = \sup_{\{\Pi_k\}} \mathcal{I}%
(\theta|\{\Pi_k\})\ ,
\end{equation}
is interpreted \cite{Zurek,Vedral} as a measure of classical correlations.
Now, these two quantities -- $\mathcal{I}(\theta)$ and $\mathcal{C}_\mathcal{%
H}(\theta)$ -- may differ and the difference
\begin{equation}
\mathcal{D}_{\mathcal{H}}(\theta) = \mathcal{I}(\theta) - \mathcal{C}_%
\mathcal{H}(\theta)
\end{equation}
is called a quantum discord.

Evidently, the above definition is not symmetric with respect to parties $%
\mathcal{H}$ and $\mathcal{K}$. However, one can easily swap the role of $%
\mathcal{H}$ and $\mathcal{K}$ to get
\begin{equation}
\mathcal{D}_{\mathcal{K}}(\theta) = \mathcal{I}(\theta) - \mathcal{C}_%
\mathcal{K}(\theta) \ ,
\end{equation}
where
\begin{equation}  \label{C-sup}
\mathcal{C}_{\mathcal{K}}(\theta) = \sup_{\{\widetilde{\Pi}_\alpha\}}
\mathcal{I}(\theta|\{\widetilde{\Pi}_\alpha\})\ ,
\end{equation}
and $\widetilde{\Pi}_\alpha$ is a collection of one-dimensional projectors
in $\mathcal{K}$ satisfying $\widetilde{\Pi}_1 + \widetilde{\Pi}_2 + \ldots
= 1_\mathcal{K}$. For a general mixed state $\mathcal{D}_\mathcal{H}(\theta)
\neq \mathcal{D}_\mathcal{K}(\theta)$. However, it turns out that $\mathcal{D%
}_\mathcal{H}(\theta),\, \mathcal{D}_\mathcal{K}(\theta) \geq 0$. Moreover,
on pure states, quantum discord coincides with the von Neumann entropy of
entanglement $S(\rho) = S(\sigma)$. States with zero quantum discord -- so
called classical-quantum states -- represent essentially a classical
probability distribution $p_k$ embedded in a quantum system. One shows that $%
\mathcal{D}_\mathcal{H}(\theta)=0$ if and only if there exists an
orthonormal basis $|k\>$ in $\mathcal{H}$ such that
\begin{equation}  \label{Q=0}
\theta = \sum_k p_k\, |k\>\<k| {\,\otimes\,} \sigma_k \ ,
\end{equation}
where $\sigma_k$ are density matrices in $\mathcal{K}$. Similarly, $\mathcal{%
D}_\mathcal{K}(\theta)=0$ if and only if there exists an orthonormal basis $%
|\alpha\>$ in $\mathcal{K}$ such that
\begin{equation}  \label{Q=0-B}
\theta = \sum_\alpha q_\alpha\, \rho_\alpha {\,\otimes\,} |\alpha\>\<\alpha|
\ ,
\end{equation}
where $\rho_\alpha$ are density matrices in $\mathcal{H}$. It is clear that
if $\mathcal{D}_\mathcal{H}(\theta)=\mathcal{D}_\mathcal{K}(\theta)=0$, then
$\theta$ is diagonal in the product basis $|k\> {\,\otimes\,} |\alpha\>$ and
hence
\begin{equation}  \label{Q=0-B}
\theta = \sum_{k,\alpha} \lambda_{k\alpha}\, |k\>\<k| {\,\otimes\,}
|\alpha\>\<\alpha| \ ,
\end{equation}
is fully encoded by the classical joint probability distribution $%
\lambda_{k\alpha}$.

Finally, let us introduce a symmetrized quantum discord
\begin{equation}  \label{D-symm}
\mathcal{D}_{\mathcal{H}:\mathcal{K}}(\theta) := \frac{1}{2} \Big[ \mathcal{D%
}_\mathcal{H}(\theta) + \mathcal{D}_\mathcal{K}(\theta) \Big] \ .
\end{equation}
Let us observe that there is an intriguing relation between (\ref{D-symm})
and (\ref{DEN}). One has
\begin{equation}
D(\theta) = I(\theta) - \frac 12 [ S(\rho) + S(\sigma) ] \ ,
\end{equation}
whereas
\begin{equation}
\mathcal{D}_{\mathcal{H}:\mathcal{K}}(\theta) = I(\theta) - \mathcal{C}_{%
\mathcal{H}:\mathcal{K}}(\theta)\ .
\end{equation}
Note, that $\mathcal{D}_{\mathcal{H}:\mathcal{K}}(\theta) \geq 0$ but $%
D(\theta)$ can be negative (for PPT states). It is assumed that $\mathcal{D}%
_{\mathcal{H}:\mathcal{K}}(\theta)$ measures perfectly quantum correlations
encoded into $\theta$.

\begin{ex}[Separable correlated state]
\label{example:separable} For the entanglement map given by
\begin{equation*}
\phi(b)= \sum_{i} \lambda_i\,\rho_i\mathrm{Tr}\sigma_i b,\quad \phi^\ast(a)
= \sum_{i} \lambda_i\,\sigma_i\mathrm{Tr}\rho_i a, \quad \biggl( %
\,\sum_i\lambda_i = 1,\,\lambda_i\geq 0\,\forall i\,\biggr),
\end{equation*}
the corresponding state $\theta$ can be written in the form
\begin{equation}
\theta = \sum_{i} \lambda_i\,\rho_i\otimes \sigma_i,
\end{equation}
with $\rho =\phi(1_\mathcal{K}) =\sum_i \lambda_i\rho_i$ and $%
\sigma=\phi^\ast(1_\mathcal{H}) = \sum_i \lambda_i\sigma_i$. Then, we have
the following inequalities.\cite{AMO2,BO1,BO2}
\begin{align}
&0\leq I(\theta) \leq \min \bigl\{S(\rho), S(\sigma)\bigr\}, \\
&S_\theta(\sigma | \rho) \geq 0,\quad S_\theta(\rho|\sigma) \geq 0.
\end{align}
\end{ex}

\begin{ex}[Separable perfectly correlated state]
Let $\{e_i\}_i$ and $\{f_j\}_j$ be the complete orthonormal systems in $%
\mathcal{H}$ and $\mathcal{K}$, respectively. For the entanglement map given
by
\begin{equation*}
\phi(b) = \sum_i \lambda_{i}\vert e_i\rangle\langle e_i\vert\langle
f_i,bf_i\rangle,\quad \phi ^\ast(a) = \sum \lambda_i\vert f_i\rangle\langle
f_i\vert\langle e_i,ae_i\rangle,
\end{equation*}
the corresponding state $\theta$ can be written in the form
\begin{equation*}
\theta = \sum \lambda_{i}\vert e_i\rangle\langle e_i\vert \otimes \vert
f_i\rangle\langle f_i\vert\ ,
\end{equation*}
with $\rho =\phi(1_\mathcal{K}) =\sum \lambda_{i}\vert e_{i}\rangle \langle
e_i\vert,\, \sigma=\phi^\ast(1_\mathcal{H}) =\sum_i\lambda_i\,\vert
f_i\rangle\langle f_i\vert$. It is clear that $\mathcal{D}_{\mathcal{H}:%
\mathcal{K}}(\theta)=0$. Moreover, one obtains
\begin{align}
I(\theta) &=S(\rho) +S(\sigma) -S(\theta_\phi) =S(\rho), \\
S_\theta(\sigma |\rho) &= S_\theta(\rho |\sigma)=0,
\end{align}
where $S(\rho)=S(\sigma)=S(\theta_\phi) =-\sum \lambda_i\log \lambda_i$.
This correlation corresponds to a perfect correlation in the classical
scheme.
\end{ex}

%%%%%%%%%%%%%%%%%%%%%%%%%%%%%%%%%%%%%%%%%%%%%%%%%%%%%%%%%%%%%%%%%%%%%%%%%

\section{Quantum correlations for circulant states}

\label{CIRC}

In this section, we analyze correlations encoded into the special family of
so called \emph{circulant states}.

\subsection{A circulant state}

We start this section by recalling the definition of a circulant state
introduced in \cite{CKcir07} (see also \cite{Art}). Consider the finite
dimensional Hilbert space $\mathbb{C}^d$ with the standard basis $%
\{e_0,\,e_1,\,\cdots,\,e_{d-1}\}$. Let $\Sigma_0$ be the subspace of $%
\mathbb{C}^d\otimes\mathbb{C}^d$ generated by $e_i\otimes
e_i~(i=0,\,1,\,\cdots,\,d-1):$

\begin{equation}
\Sigma_0 = \mathrm{span}\{e_0\otimes e_0,\,e_1\otimes e_1,\,\cdots,
e_{d-1}\otimes e_{d-1}\}.
\end{equation}
Define a shift operator $S^\alpha : \mathbb{C}^d {\, \rightarrow\, } \mathbb{%
C}^d$ by
\begin{equation*}
S^\alpha e_k = e_{k+\alpha} \ , \ \ \ \mathrm{mod}\ d
\end{equation*}
and let
\begin{equation}
\Sigma_\alpha := (1_d {\,\otimes\,} S^\alpha)\Sigma_0 \ .
\end{equation}
It turns out that $\Sigma_\alpha$ and $\Sigma_\beta~(\alpha\ne\beta)$ are
mutually orthogonal and one has the following direct sum decomposition
\begin{equation}
\mathbb{C}^d\otimes \mathbb{C}^d \cong
\Sigma_0\oplus\Sigma_1\oplus\cdots\oplus\Sigma_{d-1}.
\end{equation}
This decomposition is called a circulant decomposition.\cite{CKcir07} Let $%
a^{(0)},\,a^{(1)},\,\cdots,\,a^{(d-1)}$ be positive $d\times d$ matrices
with entries in $\mathbb{C}$ such that $\rho_\alpha$ is supported on $%
\Sigma_\alpha$. Moreover, let
\begin{equation}
\mathrm{tr}(a^{(0)}+\cdots +a^{(d-1)})=1\ .
\end{equation}
Now, for each $a^{(\alpha)} \in M_d(\mathbb{C})$ one defines a positive
operator in $\mathbb{C}^d {\,\otimes\,} \mathbb{C}^d$ be the following
formula
\begin{equation}
\vartheta_\alpha = \sum_{i,j=0}^{d-1} a^{(\alpha)}_{ij}\, e_{ij}\otimes
S^\alpha e_{ij} S^{\alpha\dagger }.
\end{equation}
Finally, let us introduce
\begin{equation}
\vartheta := \vartheta_0 \oplus \cdots \oplus \vartheta_{d-1} \ .
\end{equation}
One proves\cite{CKcir07} that $\rho$ defines a legitimate density operators
in $\mathbb{C}^d {\,\otimes\,} \mathbb{C}^d$. One calls it a \emph{circulant
state}. For further details of circulant states we refer to Refs. \cite%
{CKcir07,Art}.

Now, let consider a partial transposition of the circulant state. It turns
out that $\rho^\tau = ({%
\mathchoice{\rm 1\mskip-4mu l}{\rm 1\mskip-4mu
l}{\rm 1\mskip-4.5mu l}{\rm 1\mskip-5mu l}} {\,\otimes\,} \tau)\rho$ is
again circulant but it corresponds to another cyclic decomposition of the
original Hilbert space $\mathbb{C}^d {\,\otimes\,} \mathbb{C}^d$. Let us
introduce the following permutation $\pi$ from the symmetric group $S_d$: it
permutes elements $\{0,1,\ldots,d-1\}$ as follows
\begin{equation}
\pi(0) = 0 \ , \ \ \ \ \pi(i) = d-i \ , \ \ i=1,2,\ldots,d-1\ .
\end{equation}
We use $\pi$ to introduce
\begin{equation}
\widetilde{{\Sigma}}_0 = \mbox{span}\left\{ e_0 {\,\otimes\,} e_{\pi(0)}\, ,
e_1 {\,\otimes\,} e_{\pi(1)}\, , \ldots\, , e_{d-1} {\,\otimes\,}
e_{\pi(d-1)} \right\} \ ,
\end{equation}
and
\begin{equation}
\widetilde{{\Sigma}}_\alpha = ({\mathchoice{\rm 1\mskip-4mu l}{\rm
1\mskip-4mu l}{\rm 1\mskip-4.5mu l}{\rm 1\mskip-5mu l}} {\,\otimes\,}
S^\alpha) \widetilde{{\Sigma}}_0\ .
\end{equation}
It is clear that $\widetilde{\Sigma}_\alpha$ and $\widetilde{\Sigma}_\beta$
are mutually orthogonal (for $\alpha\neq \beta$). Moreover,
\begin{equation}  \label{D-new}
\widetilde{\Sigma}_0 \oplus \ldots \oplus \widetilde{\Sigma}_{d-1} = \mathbb{%
C}^d {\,\otimes\,} \mathbb{C}^d \ ,
\end{equation}
and hence it defines another circulant decomposition. Now, the partially
transformed state $\vartheta^\tau$ has again a circulant structure but with
respect to the new decomposition (\ref{D-new}):
\begin{equation}  \label{ro-C-new}
\vartheta^\tau = \widetilde{\vartheta}^{(0)} + \cdots + \widetilde{\vartheta}%
^{(d-1)} \ ,
\end{equation}
where
\begin{equation}
\widetilde{\vartheta}^{(\alpha)} = \sum_{i,j=0}^{d-1} \widetilde{a}%
^{(\alpha)}_{ij} \ e_{ij} {\,\otimes\,} S^\alpha e_{\pi(i)\pi(j)} S^{\dag
\alpha} \ ,\ \ \ \ \ \alpha=0,\ldots,d-1 \ ,
\end{equation}
and the new $d \times d$ matrices $[\widetilde{a}^{(\alpha)}_{ij}]$ are
given by the following formulae:
\begin{equation}  \label{a-tilde}
\widetilde{a}^{(\alpha)} \, =\, \sum_{\beta=0}^{d-1}\, a^{(\alpha+\beta)}
\circ ({\Pi} {S}^\beta)\ , \ \ \ \ \ \ \ \ \mbox{mod $d$}\ ,
\end{equation}
where ``$\circ$" denotes the Hadamard product,\footnote{%
A Hadamard (or Schur) product of two $n \times n$ matrices $A=[A_{ij}]$ and $%
B=[B_{ij}]$ is defined by
\begin{equation*}
(A \circ B)_{ij} = A_{ij} B_{ij}\ .
\end{equation*}%
} and $\Pi$ being a $d \times d$ permutation matrix corresponding to $\pi$,
i.e. $\Pi_{ij} := \delta_{i,\pi(j)}$. It is therefore clear that our
original circulant state is PPT iff all $d$ matrices $\widetilde{a}%
^{(\alpha)}$ satisfy
\begin{equation}
\widetilde{a}^{(\alpha)} \geq 0 \ , \ \ \ \ \alpha=0,\ldots,d-1\ .
\end{equation}

%%%%%%%%%%%%%%%%%%%%%%%%%%%%%%%%%

\subsection{Generalized Bell diagonal states}

The most important example of circulant states is provided by Bell diagonal
states \cite{BH1,BH2,BH3} defined by
\begin{equation}  \label{Bell}
\rho = \sum_{m,n=0}^{d-1} p_{mn} P_{mn}\ ,
\end{equation}
where $p_{mn}\geq 0$, $\ \sum_{m,n}p_{mn}=1$ and
\begin{equation}
P_{mn} = (\mathbb{I} {\,\otimes\,} U_{mn}) \,P^+_d\, (\mathbb{I} {\,\otimes\,%
} U_{mn}^\dagger)\ ,
\end{equation}
with $U_{mn}$ being the collection of $d^2$ unitary matrices defined as
follows
\begin{equation}  \label{U_mn}
U_{mn} e_k = \lambda^{mk} S^n e_k = \lambda^{mk} e_{k+n}\ ,
\end{equation}
with
\begin{equation}
\lambda= e^{2\pi i/d} \ .
\end{equation}
The matrices $U_{mn}$ define an orthonormal basis in the space $M_d(\mathbb{C%
})$ of complex $d \times d$ matrices. One easily shows
\begin{equation}
\mathrm{Tr}(U_{mn} U_{rs}^\dagger) = d\, \delta_{mr} \delta_{ns} \ .
\end{equation}
Some authors call $U_{mn}$ generalized spin matrices since
for $d=2$ they reproduce standard Pauli matrices:
\begin{equation}  \label{U-sigma}
U_{00} = \mathbb{I}\ , \ U_{01} = \sigma_1\ , \ U_{10} = i \sigma _2\ , \
U_{11} = \sigma_3\ .
\end{equation}
Let us observe that Bell diagonal states (\ref{Bell}) are circulant states
in $\mathbb{C}^d {\,\otimes\,} \mathbb{C}^d$. Indeed, maximally entangled
projectors $P_{mn}$ are supported on $\Sigma_n$, that is,
\begin{equation}  \label{Pi_n}
\Pi_n = P_{0n} + \ldots + P_{d-1,n} \ ,
\end{equation}
defines a projector onto $\Sigma_n$, i.e.
\begin{equation}
\Sigma_n = \Pi_n ( \mathbb{C}^d {\,\otimes\,} \mathbb{C}^d) \ .
\end{equation}
One easily shows that the corresponding matrices $a^{(n)}$ are given by
\begin{equation}
a^{(n)}= H D^{(n)} H^* \ ,
\end{equation}
where $H$ is a unitary $d\times d$ matrix defined by
\begin{equation}
H_{kl} := \frac{1}{\sqrt{d}}\, \lambda^{kl} \ ,
\end{equation}
and $D^{(n)}$ is a collection of diagonal matrices defined by
\begin{equation}
D^{(n)}_{kl} := p_{kn} \delta_{kl}\ .
\end{equation}
One has
\begin{equation}
a^{(n)}_{kl} = \frac 1d \sum_{m=0}^{d-1} p_{mn} \lambda^{m(k-l)}\ ,
\end{equation}
and hence it defines a circulant matrix
\begin{equation}
a^{(n)}_{kl} = f^{(n)}_{k-l}\ ,
\end{equation}
where the vector $f^{(n)}_m$ is the inverse of the discrete Fourier
transform of $p_{mn}$ ($n$ is fixed).

%%%%%%%%%%%%%%%%%%%%%%%%%%%%%%%%%%%%%%

\subsection{A family of Horodecki states}

Let $\mathcal{H}=\mathcal{K}= \mathbb{C}^{3}$. For any $\alpha \in \lbrack
0,5]$, one defines\cite{HHHmix01} the following state
\begin{align}
\theta _{1}(\alpha ) & = \frac{2}{7}\, P^+_3 + \frac\alpha 7\, \Pi_1 + \frac{%
5-\alpha }{7} \, \Pi_2\ .
\end{align}
The eigenvalues of $\theta _{1}(\alpha )$ are calculated as $0,\frac{2}{7}%
,\,3 \times \frac{\alpha }{21}$ and $3 \times \frac{5-\alpha }{21}$ and
hence one obtains for the $D$-correlations
\begin{equation}
D\bigl(\theta _{1}(\alpha )\bigr)= \log 3 + \frac{2}{7}\log \frac{2}{7} +
\frac{\alpha }{7}\log \frac{\alpha }{21} + \frac{5-\alpha }{7}\log \frac{%
5-\alpha }{21}\ .
\end{equation}

\begin{thm}
\label{section5:thm1} \cite{HHHmix01} The family $\theta_1(\alpha)$
satisfies:

\begin{enumerate}

\item $\theta_1(\alpha)$ is PPT if and only $\alpha \in [1,4]$

\item $\theta_1(\alpha)$ is separable if and only if $\alpha \in [2,3]$%
\textrm{;}

\item $\theta_1(\alpha)$ is both entangled and PPT if and only if $\alpha
\in [1,2) \cup (3,4]$ \textrm{;}

\item $\theta_1(\alpha)$ is NPT if and only if $\alpha \in [0,1) \cup (4,5]$.
\end{enumerate}
\end{thm}

Due to this Theorem, one can find that the $D(\theta _{1}(\alpha ))$ does
admit a natural order. That is, the $D$-correlation for any entangled state
is always stronger than $D$-correlation for an arbitrary separable state.
Similarly, one observes that $D$-correlation for any NPT state is always
stronger than $D$-correlation for an arbitrary PPT state. The graph of $D%
\bigl(\theta _{1}(\alpha )\bigr)$ is shown in Fig.~\ref{section5:fig1}.
Actually, one finds that the minimal value of $D$-correlations corresponds
to $\alpha = 2.5$, that is, it lies in the middle of the separable region.

On the other hand, we can also compute the symmetrized discord $\mathcal{D}_{%
\mathbb{C}^{3};\mathbb{C}^{3}}\left( \theta _{1}\left(\alpha \right) \right)$
and have obtained Fig.~\ref{section5:fig1}.
%( We have collected all sorts of numerical data of
%$\mathcal{D}_{\mathbb{C}^{3};\mathbb{C}^{3}}\left( \theta _{1}\left( \alpha \right) \right)$. See Appendix A.).
It is easy to find that the graph is symmetric with respect to $\alpha =2.5$%
. As in Fig. 2, the value of the symmetrized discord satisfies the following
inequality;
\begin{equation*}
0 < \mathcal{D}_{\mathbb{C}^{3};\mathbb{C}^{3}}\left(\theta
_{1}(\alpha)\right) \,\leq\, \mathcal{D}_{\mathbb{C}^{3};\mathbb{C}%
^{3}}\left(\theta _{1}(\beta)\right) \,\leq\, \mathcal{D}_{\mathbb{C}^{3};%
\mathbb{C}^{3}}\left(\theta _{1}(\gamma)\right),
\end{equation*}
where $\alpha \in \left[ 2,3\right],\,\beta \in \left[1,2\right] \cup \left[%
3,4\right]$ and $\gamma \in \left[0,1\right] \cup \left[4,5\right]$.

The family of $\theta _{1}\left( \alpha \right) $ has the quantum
correlation even in separable states corresponding to $\alpha \in \left[ 2,3%
\right] $ in the sense of discord. We know that the above two types of
criteria give the similar order of correlation.

Notice that $D\left( \theta _{1}\left( \alpha \right) \right) $ is always
negative even in NPT sates and the positivity of $D$-correlation represents
a true quantum property (see Example 3.3 and Proposition 3.5). In this sense
the quantum correlation of $\theta _{1}\left( \alpha \right) $ is not so
strong.

\begin{figure}[tbph]
\begin{center}
\includegraphics[width=6.0cm]{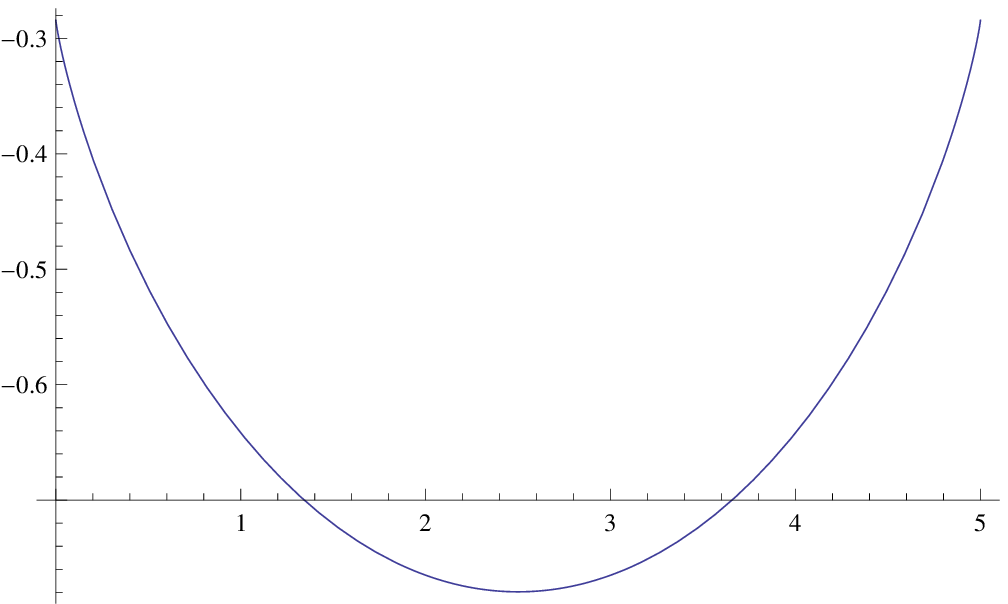} \hspace{1cm} %
\includegraphics[width=6.0cm]{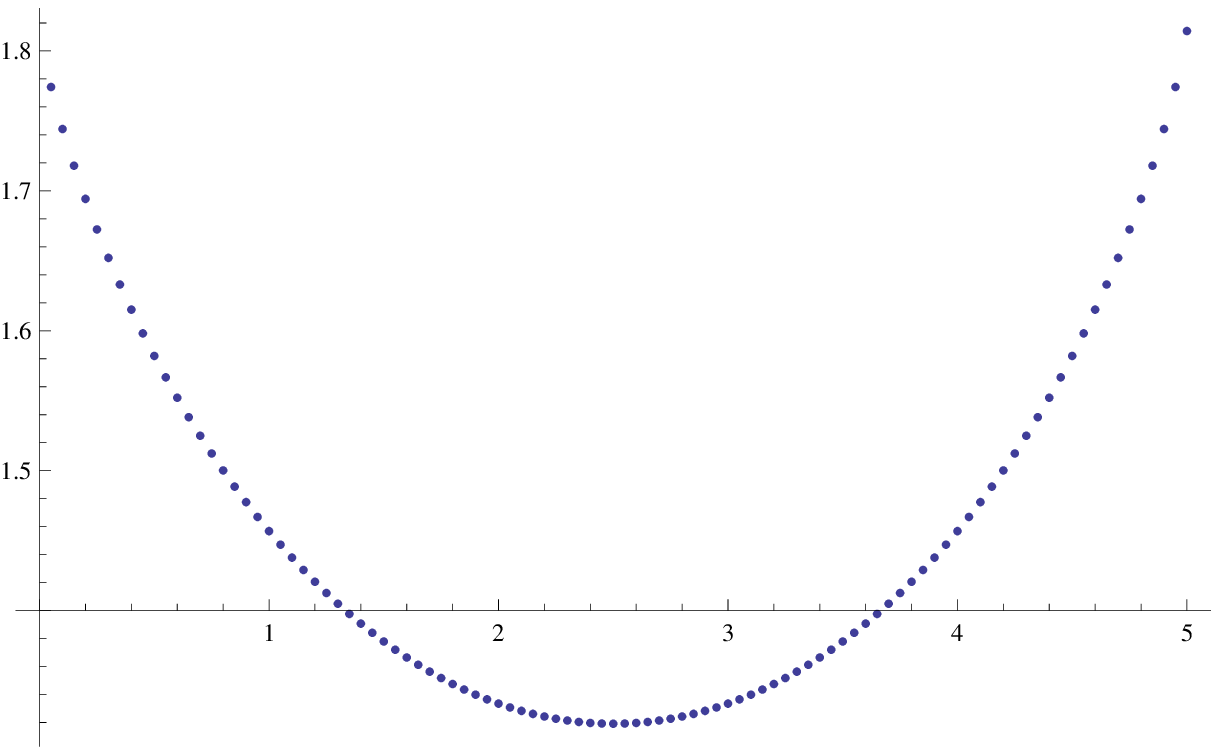}
\end{center}
\caption{Left --- the graph of $D(\protect\theta _{1}(x))\,$ with $x \in
[0,5]$. The minimal value of $D$ corresponds to $x=2.5$. Right --- the graph
of $\mathcal{D}_{\mathbb{C}^{3};\mathbb{C}^{3}}\left( \protect\theta %
_{1}\left( \protect\alpha \right) \right)$. }
\label{section5:fig1}
\end{figure}

This family may be generalized to $\mathbb{C}^d {\,\otimes\,} \mathbb{C}^d$
as follows: consider the following family of circulat 2-qudit states
\begin{equation}  \label{rho}
\theta(\alpha) = \sum_{i=1}^{d-1} \lambda_i \Pi_i + \lambda_d P^+_d\ ,
\end{equation}
with $\lambda_n\geq 0$, and $\lambda_1 + \ldots +\lambda_{d-1} + \lambda_d=1$%
. Let us take the following special case corresponding to
\begin{eqnarray}  \label{lll}
\lambda_1 = \frac{ \alpha}{\ell} \ , \ \ \lambda_{d-1} = \frac{(d-1)^2+1
-\alpha}{\ell} \ , \ \ \lambda_d = \frac{d-1}{\ell} \ .
\end{eqnarray}
and $\lambda_2 = \ldots = \lambda_{d-2} = \lambda_d$, with
\begin{equation}
\ell = (d-1)(2d-3) +1\ .
\end{equation}
One may prove the following\cite{Adam}

\begin{thm}
The family $\theta(\alpha)$ satisfies:

\begin{enumerate}
\item $\theta(\alpha)$ is PPT if and only $\alpha \in [1,(d-1)^2]$

\item $\theta(\alpha)$ is separable if and only if $\alpha \in
[d-1,(d-1)(d-2)+1]$\textrm{;}

\item $\theta_1(\alpha)$ is both entangled and PPT if and only if $\alpha
\in [1,d-1) \cup ((d-1)(d-2)+1,(d-1)^2]$ \textrm{;}

\item $\theta_1(\alpha)$ is NPT if and only if $\alpha \in [0,1) \cup
((d-1)^2,(d-1)^2+1]$.
\end{enumerate}
\end{thm}

For example if $d=4$ one obtains the following picture of $D(\theta(\alpha))$
(see Fig.~4)
\begin{figure}[tbph]
\begin{center}
\includegraphics[width=6.0cm]{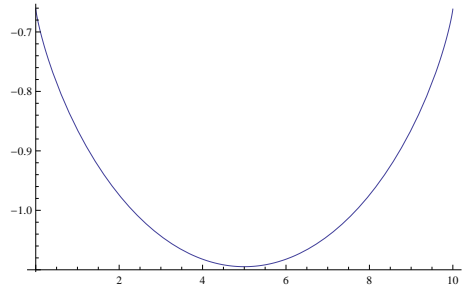}
\end{center}
\caption{The graph of $D(\protect\theta(x))\,$ with $x \in [0,10]$. The
minimal value of $D$ corresponds to $x=5$.}
\label{section5:fig1}
\end{figure}
Again, one finds that the $D(\theta(\alpha ))$ does admit a natural order.
That is, the $D$-correlation for any entangled state is always stronger than
$D$-correlation for an arbitrary separable state. Similarly, one observes
that $D$-correlation for any NPT state is always stronger than $D$%
-correlation for an arbitrary PPT state.

%%%%%%%%%%%%%%%%%%%%%%%%%%%%%%%%%%%%%%%

\subsection{Example: a family of Bell diagonal states}

Consider the following class of Bell-diagonal states in $\mathbb{C}^3 {%
\,\otimes\,} \mathbb{C}^3$:
\begin{equation}
\theta_2(\varepsilon) = \frac{1}{\Lambda } ( 3P^+_3 + \varepsilon \Pi_1 +
\varepsilon^{-1} \Pi_2 ) \ ,
\end{equation}
with $\Lambda =1+\varepsilon +\varepsilon ^{-1}$. One easily finds for its $%
D $-correlations
\begin{eqnarray}
D\bigl(\theta _{2}(\varepsilon )\bigr) = \frac 1\Lambda \left( \log \frac
1\Lambda + \varepsilon^{-1} \log \frac{\varepsilon^{-1}}{\Lambda} +
\varepsilon \log \frac{\varepsilon}{\Lambda} + \log 3 \right) \ .
\end{eqnarray}

The following theorem gives us a useful characterization of $%
\theta_{2}(\varepsilon )$ \cite{JCRacl09}.

\begin{thm}
\label{section3:thm2} The states of $\theta_1(\varepsilon)$ are classified
by $\varepsilon$ as follows:

\begin{enumerate}

\item $\theta_2(\varepsilon)$ is separable if $\varepsilon =1$\textrm{;}

\item $\theta_2(\varepsilon)$ is both PPT and entangled for $\varepsilon
\neq 1$.
\end{enumerate}
\end{thm}

\begin{figure}[tbph]
\begin{center}
\includegraphics[width=6.0cm]{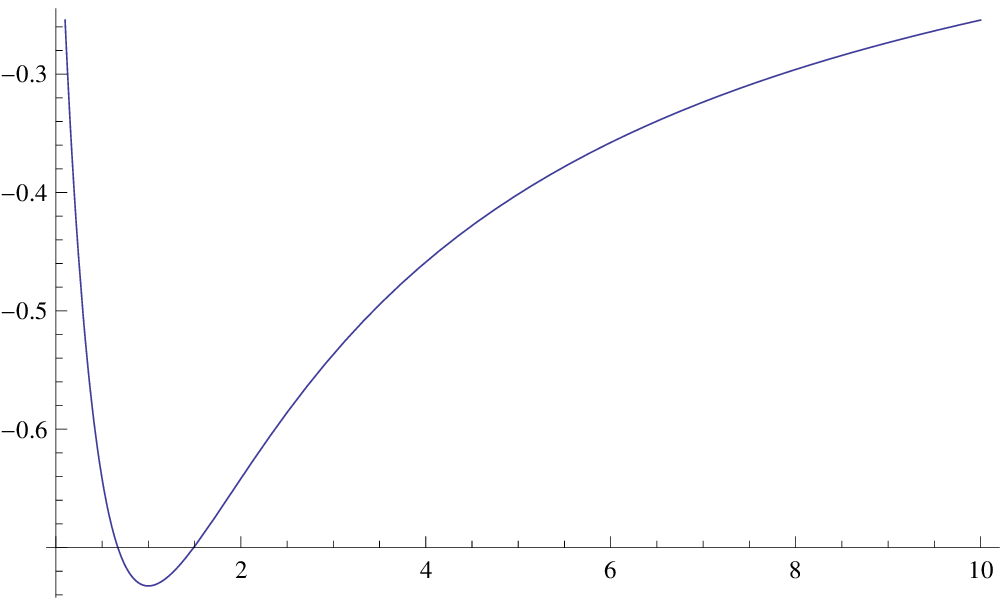} \hspace{1cm} %
\includegraphics[width=6.0cm]{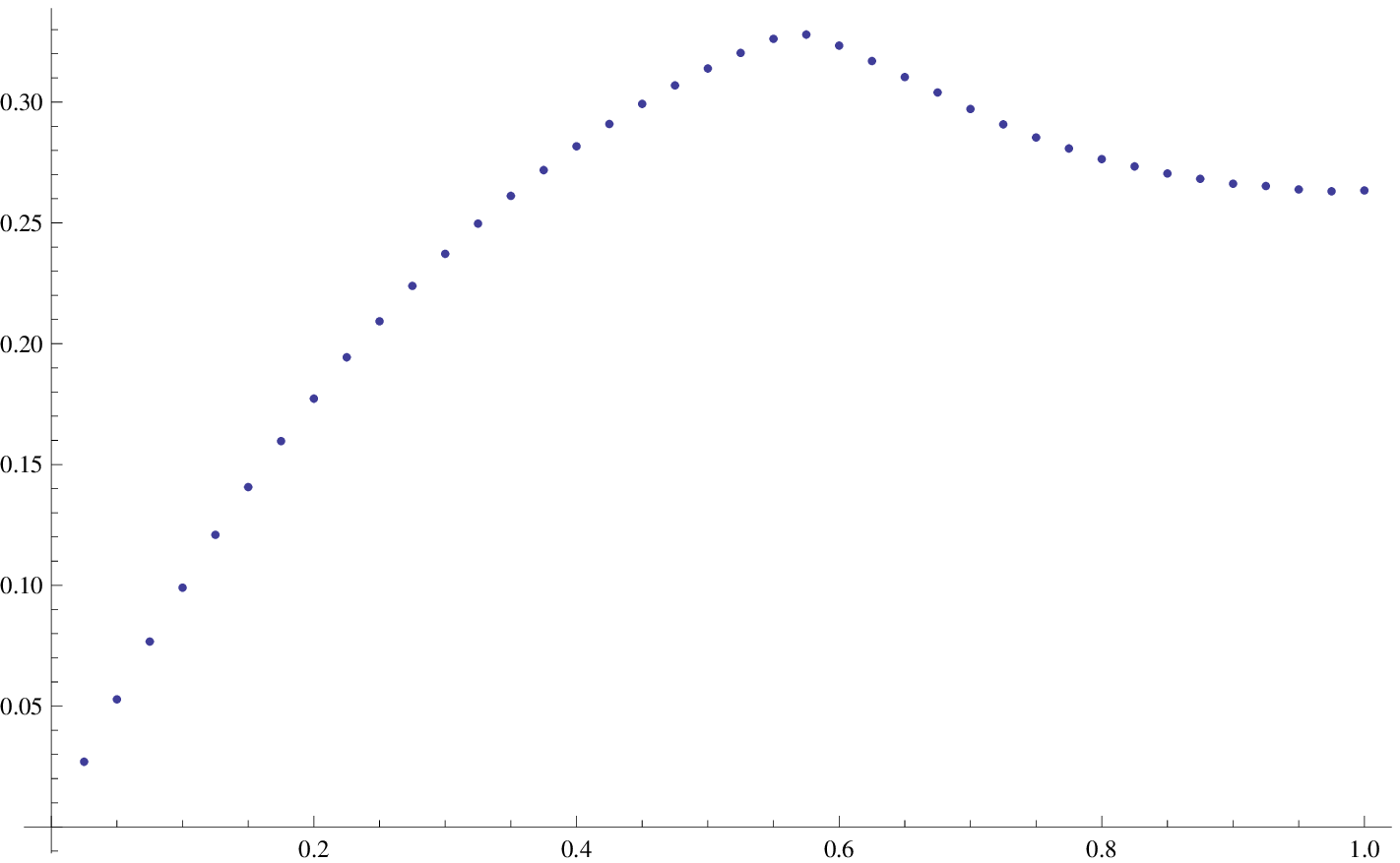}
\end{center}
\caption{Left --- the graph of $D(\protect\theta _{2}(x))$. Note that $D$ is minimal for $x=1$ which correspond to the separable state.
Right --- the graph of $\mathcal{D}_{\mathbb{C}^3:\mathbb{C}^3}(\theta_2(\varepsilon)$ for $\varepsilon\in (0,1]$. Note that $\,\mathcal{D}_{\mathbb{C}^3:\mathbb{C}^3}(\theta_2(\varepsilon))= \mathcal{D}_{\mathbb{C}^3:\mathbb{C}^3}(\theta_2(\varepsilon^{-1}))$. }
\label{section3:fig2}  
\end{figure}

\noindent The graph of $D\bigl(\theta _{2}(\varepsilon )\bigr)$ is shown in
Fig.~\ref{section3:fig2}. $D\bigl(\theta _{2}(\varepsilon )\bigr)$ is
rapidly decreasing with $\varepsilon$ approaching $1$ from $0$ and increases
when $\varepsilon$ is over $1$. That is, $D\bigl(\theta _{2}(\varepsilon )%
\bigr)$ takes the minimal value at $\varepsilon =1$ and it is approximated
about $D\bigl(\theta _{2}(1)\bigr)=-\frac{2}{3}\log 3\approx -0.7324$. As is
the case of $\theta_1(\alpha)$, the $D$-correlation $D\bigl(\theta
_{2}(\varepsilon )\bigr)$ for an entangled state is always stronger than the
one for a separable state. As $\varepsilon\rightarrow 0$ or $\infty $, $%
\theta _{2}\left( \varepsilon \right) $ converges to a separable perfectly
correlated state which can be recognized as a \textquotedblleft classical
state\textquotedblright\

\begin{equation}
\lim_{\varepsilon \rightarrow 0 }\theta _{2}(\varepsilon )=\frac{1}{3}\Bigl(%
e_{00}\otimes e_{22}+e_{11}\otimes e_{00}+e_{22}\otimes e_{11}\Bigr) =
\Pi_2\ ,  \label{section3:eqn2}
\end{equation}

\begin{equation}
\lim_{\varepsilon \rightarrow \infty}\theta _{2}(\varepsilon )= \frac{1}{3}%
\Bigl(e_{00}\otimes e_{11}+e_{11}\otimes e_{22}+e_{22}\otimes e_{00}\Bigr) =
\Pi_1\ ,
\end{equation}
and for every $\varepsilon >0$,
\begin{equation}
D\bigl(\theta _{2}(\varepsilon )\bigr)<0=\lim_{\varepsilon \rightarrow 0}D%
\bigl(\theta _{2}(\varepsilon )\bigr)=\lim_{\varepsilon \rightarrow \infty }D%
\bigl(\theta _{2}(\varepsilon )\bigr).
\end{equation}%
It shows that a correlation of a PPT entangled state $\theta _{2}\left(
\varepsilon \neq 1\right) $ is weaker than that of the (classical) separable
perfectly correlated states in the sense of (\ref{order}).

Now, since $\theta _{1}(\alpha )$ and $\theta _{2}(\varepsilon )$ have
common marginal states, we can compare the order of quantum correlations for
them. One has, for example,
\begin{equation}
D\bigl(\theta _{2}(1)\bigr)\approx -0.7324>-0.7587\approx D\bigl(\theta
_{1}(3.1)\bigr).  \label{section3:ineq1}
\end{equation}%
Accordingly Theorem \ref{section5:thm1} and \ref{section3:thm2}, however, $%
\theta _{2}(1)$ is separable while $\theta _{1}(3.1)$ is entangled state.
Incidentally, this means that the correlation for the separable state $%
\theta _{2}(1)$ is stronger than the entangled state $\theta _{1}(3.1)$ in
the sense of (\ref{order}).

On the other hand one finds the following plot of the quantum discord Fig.~3.

It is clear that
\begin{equation}
\lim_{\varepsilon \rightarrow 0} \mathcal{D}_{\mathbb{C}^3:\mathbb{C}%
^3}(\theta_2(\varepsilon)) = \lim_{\varepsilon \rightarrow \infty}\mathcal{D}%
_{\mathbb{C}^3:\mathbb{C}^3}(\theta_2(\varepsilon)) = 0 \ ,
\end{equation}
since both $\Pi_1$ and $\Pi_2$ are perfectly classical states. Note, that $%
\mathcal{D}_{\mathbb{C}^3:\mathbb{C}^3}(\theta_2(\varepsilon=1))> 0$ which
shows that separable state $\theta_2(\varepsilon=1)$ does contain quantum
correlations.

%%%%%%%%%%%%%%%%%%%%%%%%%%%%%%%%%%%%%%%%%%%%%%%%%%%%%%%%%%%%%%%%%%%%%%%%%

%%%%%%%%%%%%%%%%%%%%%%%%%%%%%%%%%%%%%%%%%%%%%%%%%%%%%%%%%%%%%%%%%%%%%%%%%

\section{Conclusions}

We provided several examples of bi-partite quantum states and computed two
types of correlations for them. It turned out that the correlation for a
separable state can be stronger than the one for an entangled state in the
sense of (\ref{order}). This observation is inconsistent with the
conventional understanding of quantum entanglement. However, we also showed
that the discord of such separable states might strictly positive. This
means that these states have a non-classical correlation. From this point of
view, it is no longer unusual that the correlation for a separable state is
stronger than the one for an entangled state.

\section*{Acknowledgments}

T.M. is grateful to V.P.Belavkin for fruitful discussions on mutual entropy
and entanglement maps. D.C. was partially supported by the National Science
Centre project DEC-2011/03/B/ST2/00136. We would like to acknowledge the
supports of QBIC (Quantum Bio-Informatics Center) grant of Tokyo University
of Science.

%%%%%%%%%%%%%%%%%%%%%%%%%%%%%%%%%%%%%%%%%%%%%%%%%%%%%%%%%%%%%%%%%%%%%%


\begin{thebibliography}{99}
\bibitem{ACKMO} L.Accardi, D. Chru\'{s}ci\'{n}ski, A. Kossakowski, T.
Matsuoka and M. Ohya, "On classical and quantum liftings", Open. Syst. Info.
Dyn. \textbf{17}, 361--386 (2010).

\bibitem{AMO} L. Accardi, T. Matsuoka, M. Ohya, "Entangled Markov chains are
indeed entangled", Infin. Dim. Anal. Quantum Probab. Top. \textbf{9},
379--390 (2006).

\bibitem{AMO2} L.Accardi, T. Matsuoka, M. Ohya,"Entangled Markov chain
satisfying entanglement condition", RIMS \textbf{1658}, 84--94 (2009).

\bibitem{AO} L.Accardi, M. Ohya,"Composite channels, transition expectation
and liftings", J. Appl. Math. Optim., \textbf{39}, 33--59 (1999).

\bibitem{AKMS} M. Asorey, A. Kossakowski, G. Marmo, E.C.G. Sudarshan,
"Relation between quantum maps and quantum state", Open. Syst. Info. Dyn.
\textbf{12}, 319--329 (2006).

\bibitem{BH1} B. Baumgartner, B. Hiesmayer and H. Narnhofer, "State space
for two qutrits has a phase space structure in its cone", Phys. Rev. A
\textbf{74}, 032327 (2006).

\bibitem{BH2} B. Baumgartner, B. Hiesmayer and H. Narnhofer, "A special
simplex in the state space for entangled qutrits", J. Phys. A: Math. Theor.,
\textbf{40}, 7919 (2007).

\bibitem{BH3} B. Baumgartner, B. Hiesmayer and H. Narnhofer, "The geometry
of biparticle qutrits including bound entanglement", Phys. Lett. A, \textbf{%
\ 372}, 2190 (2008).

\bibitem{Be1} V. P. Belavkin, "Optimal filtering of Markov signals with
quantum white noise", Radio Eng. Electron. Phys., \textbf{25}, 1445-1453
(1980).

\bibitem{Be2} V. P. Belavkin, "Nondemolition principle of quantum
measurement theory", Found. Phys., \textbf{24}, 685--714 (1994).

\bibitem{BO1} V. P. Belavkin, M. Ohya, "Quantum entropy and information in
discrete entangled state", Infin. Dim. Anal. Quantum Probab. Top. \textbf{4}%
, 33-59 (2001).

\bibitem{BO2} V. P. Belavkin, M. Ohya, "Entanglement, quantum entropy and
mutual information", Proc. R. Soc. London A \textbf{458}, 209--231 (2002).

\bibitem{BD} V. P. Belavkin, X. Dai, "An operational algebraic approach to
quantum channel capacity", Int. J. Quantum Inf. \textbf{6}, 981 (2008).

\bibitem{Cerf} N. J. Cerf and C. Adami, "Negative entropy and information in
quantum mechanics", Phys. Rev. Lett. \textbf{79}, 5194--5197 (1997).

\bibitem{Ch} M. D. Choi, "Completely positive maps on complex matrix", Lin.
Alg. Appl., \textbf{10}, 285 (1975).

\bibitem{CHMO} D. Chru\'{s}ci\'{n}ski, Y. Hirota, T. Matsuoka and M. Ohya,
"Remarks on the degree of entanglement" QP- PQ Quantum Probab. \& White
Noise Anal. \textbf{28}, 145--156 (2011).

\bibitem{CKcir07} D. Chru$\mathrm{\acute{s}}$ci$\mathrm{\acute{n}}$ski and
A. Kossakowski: ``Circulant states with positive partial transpose'',
Physical Review A 76 (2007), 14 pp.

\bibitem{Art} D. Chru\'sci\'nski and A. Pittenger, \textit{Generalized
Circulant Densities and a Sufficient Condition for Separability}, J. Phys.
A: Math. Theor. \textbf{41} (2008) 385301.

\bibitem{CKMM} D. Chru\'{s}ci\'{n}ski, A. Kossakowski, T. Matsuoka, K. M\l %
odawski, "A class of Bell diagonal states and entanglement witnesses", Open.
Syst. Info. Dyn. \textbf{17}, 213--231 (2010).

\bibitem{CKMO} D. Chru\'{s}ci\'{n}ski, A. Kossakowski, T. Matsuoka and M.
Ohya, "Entanglement mapping vs. quantum conditional probability operator",
QP- PQ Quantum Probab. \& White Noise Anal. \textbf{28}, 223--236 (2011).

\bibitem{Adam} D. Chru\'sci\'nski and A. Rutkowski, \textit{Entanglement
witnesses for $d \otimes d$ systems and new classes of entangled qudit states%
}, Eur. Phys. J. D \textbf{62}, 273 (2011).

\bibitem{Gro} B. Groisman, S. Popescu and A. Winter, "Quantum, classical,
and total amount of correlations in a quantum state" Phys. Rev A \textbf{72}
, 0323187 (2005).

\bibitem{Vedral} L. Henderson and V. Vedral, "Classical, quantum and total
correlation", J. Phys. A \textbf{34} 6913 (2001)

\bibitem{HH} M. Horodecki and R. Horodecki, "Information-theoretical aspect
of quantum inseparability of mixed states", Phys. Rev. A \textbf{54},
1838-1843 ( 996).

\bibitem{HHH} M. Horodecki, P. Horodecki and R. Horodecki, "Separability of
mixed states: necessary and sufficient condition", \textit{Phy. Lett. },
\textbf{A} \textbf{223}, 1 (1996).

\bibitem{HHHmix01} M. Horodecki, P. Horodecki and R. Horodecki: ``Mixed
state entanglement and quantum condition'', In Quantum Information, Springer
Tracts in Modern Physics 173 (2001), pp 151--195.

\bibitem{HHHH} R. Horodecki, P. Horodecki, M. Horodecki, and K. Horodecki,
"Quantum entanglement", Rev. Mod. Phys. \textbf{81}, 865 (2009).

\bibitem{Ja} A. Jamio\l kowski, "Linear transformation which preserve trace
and positive semidefiniteness of operators", Rep. Math. Phys. \textbf{3},
275 (1072).

\bibitem{JMO} A. Jamio\l kowski, T. Matsuoka and M. Ohya, "Entangling
operator and PPT condition", TUS preprint (2007).

\bibitem{JCRacl09} J. Jurkowski, D. Chru$\mathrm{\acute{s}}$ci$\mathrm{%
\acute{n}}$ski and A. Rutkowski: ``A class of bound entangled states of two
qutrits'', Open Syst. Inf. Dyn., 16 (2009), no 2-3, pp 235--242.

\bibitem{KK} G. Kimura, A. Kossakowski, "A note on positive maps and
classification on states", Open Sys. \& Information Dyn. \textbf{12},
1(2005).

\bibitem{MMO} W. A. Majewski, T. Matsuoka and M. Ohya, "Characterization of
partial positive transposition states and measures of entanglement", J.
Math. Phys. \textbf{50}, 113509 (2009).

\bibitem{Ma} T. Matsuoka, "On generalized entanglement", QP- PQ Quantum
Probab. \& White Noise Anal. \textbf{21}, 170--180 (2007).

\bibitem{MO} T. Matsuoka, M. Ohya, "Quantum entangled state and its
characterization", Foud. Probab. Phys. 3 \textbf{750}, 298--306 (2005).

\bibitem{MP} K. Modi, T. Paterek, W. Son, V. Vedral and M. Williamson,
"Unified view of quantum and classical correlations", Phys. Rev. Lett.,
\textbf{104}, 080501 (2010).

\bibitem{Oh1} M. Ohya, "On composite state and mutual information in quantum
information theory", IEEE Info. Theory, \textbf{29}, 77--774 (1983).

\bibitem{Oh2} M. Ohya, "Note on quantum probability", Nuovo Cimento, \textbf{%
38}, 402-406 (1983).

\bibitem{OV} M. Ohya, I. V. Volovich, \textit{Mathematical Foundations of
Quantum Information and Computation and Its Applications to Nano- and
Bio-systems}, Springer, New York, 2011.

\bibitem{Zurek} H. Ollivier, W. Z. Zurek, "Quantum discord: A measure of the
quantumness of correlations", Phys. Rev. Lett., \textbf{88}, 017901 (2002).

\bibitem{Paulsen} V. Paulsen, \emph{Completely Bounded Maps and Operator
Algebras}, Cambridge University Press, 2003.

\bibitem{Pe} A. Peres, "Separability criterion for density matrix", \textit{%
Phys.Rev.Lett.}, \textbf{77, }1413 (1996).

\bibitem{Sh} C. E. Shannon, "A mathematical theory of communication",
Urbana, IL: Univ. Illinois (1949).

\bibitem{Ume} H. Umegaki, "Conditional expectation in an operator algebras
IV", Kodai Math. Sem. Rep., \textbf{14}, 59--85 (1962).

%\bibitem{Ur} K. Urbanik, "Joint probability distribution of observables in
%quantum mechanics", Stud. Math. T. \textbf{21}, 317--323 (1961).

\bibitem{VW} K.G.H. Vollbrecht, M.M. Wolf, "Conditional entropies and their
relation to entanglement criteria", J. Math. Phys. \textbf{43}, 4299 (2002).

\bibitem{Wo} S. L. Woronowicz, "Positive maps of low dimensional matrix
algebra", Rep. Math. Phys., \textbf{10}, 165 (1976)
\end{thebibliography}
\end{document}